\documentclass[letterpaper]{article} 
\usepackage{aaai25}  
\usepackage{times}  
\usepackage{helvet} 
\usepackage{courier} 
\usepackage[hyphens]{url}  
\usepackage{graphicx} 
\usepackage{natbib}  
\usepackage{caption} 
\usepackage{algorithm}
\usepackage{algorithmic}

\usepackage{amsmath}
\usepackage{amssymb}

\usepackage{newfloat}
\usepackage{listings}
\DeclareCaptionStyle{ruled}{labelfont=normalfont,labelsep=colon,strut=off} 
\floatstyle{ruled}
\newfloat{listing}{tb}{lst}{}
\floatname{listing}{Listing}

\nocopyright
\title{Assessing Human Intelligence Augmentation Strategies Using Brain Machine Interfaces and Brain Organoids in the Era of AI Advancement}
\author{
    Kenta Kitamura\textsuperscript{\rm 1}
}
\affiliations{
    \textsuperscript{\rm 1}Intellectual Monnshine\\
    kitamura.kenta.1988@gmail.com
}

\usepackage{bibentry}

\begin{document}

\maketitle

\begin{abstract}
The rapid advancement of Artificial Intelligence (AI) technologies, including the potential emergence of Artificial General Intelligence (AGI) and Artificial Superintelligence (ASI), has raised concerns about AI surpassing human cognitive capabilities. To address this challenge, intelligence augmentation approaches, such as Brain Machine Interfaces (BMI) and Brain Organoid (BO) integration have been proposed. In this study, we compare three intelligence augmentation strategies, namely BMI, BO, and a hybrid approach combining both. These strategies are evaluated from three key perspectives that influence user decisions in selecting an augmentation method: information processing capacity, identity risk, and consent authenticity risk. First, we model these strategies and assess them across the three perspectives. The results reveal that while BO poses identity risks and BMI has limitations in consent authenticity capacity, the hybrid approach mitigates these weaknesses by striking a balance between the two. Second, we investigate how users might choose among these intelligence augmentation strategies in the context of evolving AI capabilities over time. As the result, we find that BMI augmentation alone is insufficient to compete with advanced AI, and while BO augmentation offers scalability, BO increases identity risks as the scale grows. Moreover, the hybrid approach provides a balanced solution by adapting to AI advancements. This study provides a novel framework for human capability augmentation in the era of advancing AI and serves as a guideline for adapting to AI development.
\end{abstract}

\section{Introduction}

\subsection{Background}

The rapid evolution of Artificial Intelligence (AI) technologies has heightened the prospect of Artificial General Intelligence (AGI) and Artificial Superintelligence (ASI), potentially surpassing human cognitive abilities and raising significant safety and ethical concerns \cite{Chalmers2016Singularity, Kurzweil2005Singularity}. In response, transhumanist and posthumanist approaches advocate for intelligence augmentation through emerging technologies such as Brain Machine Interfaces (BMI), genetic modifications, and Brain Organoid (BO) \cite{EnhanceAndIdentity, TranshumanistMedicine, Cyborg}.

Among the augmentation methods, BMI is expected to enable bidirectional information flow between the brain and external devices, offering promising opportunities for intelligence augmentation \cite{BMIReview, BMIConcern}. On the other hand, BO enhances cognitive abilities by integrating brain tissue derived from stem cells into the brain \cite{BOReview, BOConcern}.

The approach of adding BMI and BO to the brain is considered a useful countermeasure against AI.  
Intelligence augmentation approaches using BMI and BO are generally categorized into brain replacement approaches and brain add-on approaches \cite{BMIReview, BMIConcern, BOReview, BOConcern}.  
The replacement approach faces significant limitations due to the constraints of the skull and its relatively limited scalability in adapting to AI advancements.  
In contrast, the brain add-on approach provides a more scalable augmentation solution that is less affected by the inherent limitations of the brain, making it a more viable strategy to counter the progress of AI.

\begin{figure}[t]
\begin{center}
\includegraphics[width=1\linewidth]{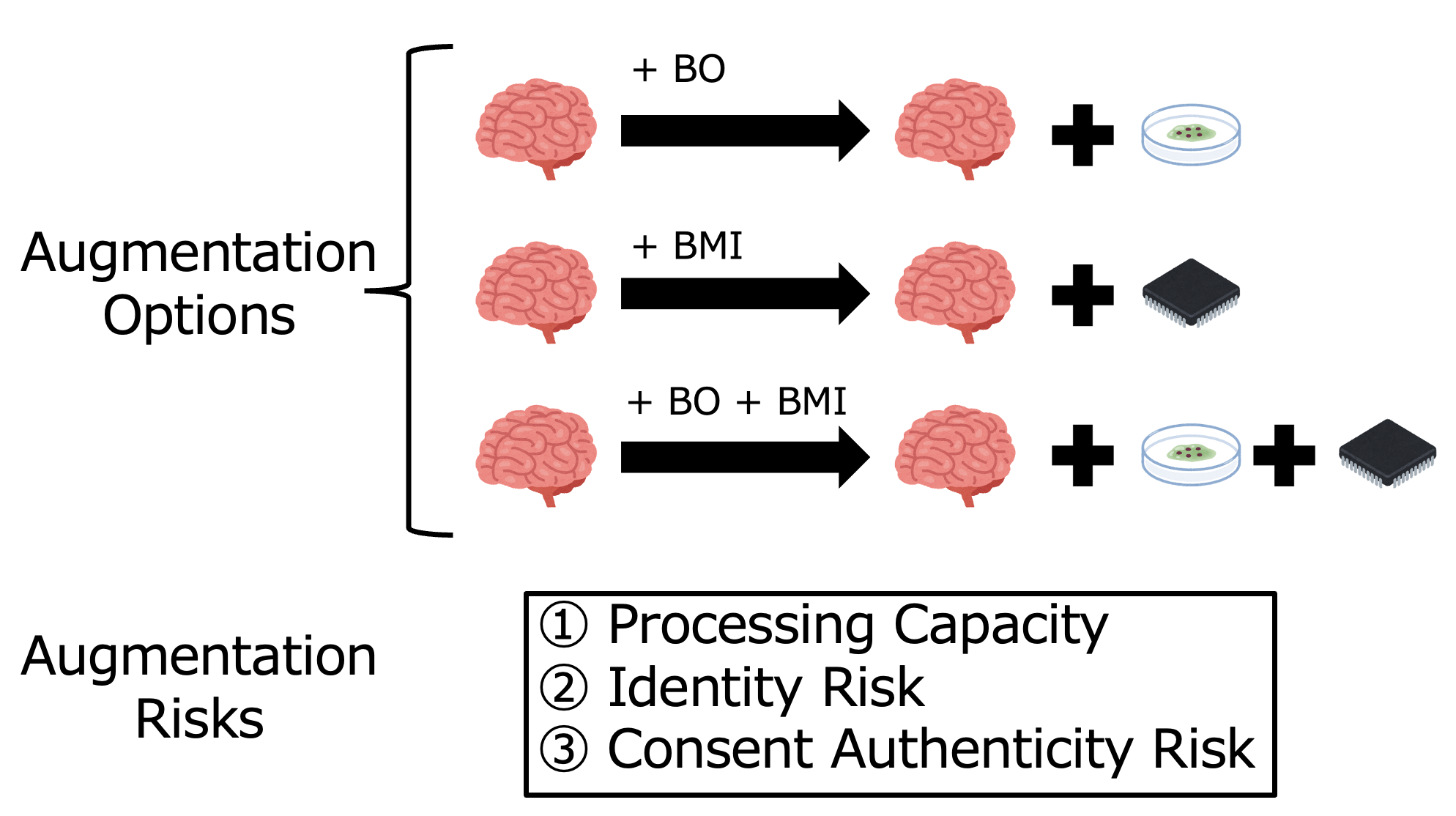}
\end{center}
\caption{Overview of augmentation approaches and risks.}
\label{fig::Overview}
\end{figure}

While BMI and BO hold promising prospects, they are also associated with certain risks.
BMI raises concerns about the exacerbation of social inequalities and security risks \cite{BMIReview, BMIConcern}. On the other hand, BO is subject to ethical concerns \cite{BOReview, BOConcern}.
In particular, when users consider these technologies for intelligence augmentation, BMI may compromise the consent authenticity, and BO poses identity risks.

\subsection{Our Contribution}

As shown in Figure~\ref{fig::Overview}, in this study, we investigate the potential risks of adding BMI, BO, and hybrid systems combining both technologies \cite{HybierdSystems} to the brain as human augmentation solutions in the context of AI advancements.
In particular, in this study, we examine the risks associated with BMI, BO, or their combination for users considering these augmentation methods.

Various risks have been proposed regarding intelligence augmentation using BMI and BO \cite{BMIReview, BMIConcern, BOConcern, BOReview}.
Among these risks, we focus on evaluating these technologies based on three key perspectives relevant to users selecting methods for intelligence augmentation to counter AI advancements. These perspectives are the scalability of information processing capacity, the personal identity risk, and the consent authenticity risk.

First, we model the brain and the three approaches. Then, we evaluate the model using the three perspectives described above. The results indicate that while BO poses identity risks and BMI faces limitations in consent authenticity, the hybrid approach mitigates these weaknesses.

Second, using the framework, we illustrate how each technology can be applied as AI advances. The evaluation results indicate that BMI-based augmentation alone may reach its limits in countering AI, necessitating a transition to BO or hybrid systems.

In summary, our contributions are as follows:

\begin{enumerate}
    \item We propose a model for evaluating intelligence augmentation technologies in terms of scalability of information processing capacity, self-identity risk, and consent authenticity risk to counter the advancement of AI.
    \item Based on the model, we conduct a scenario analysis to assess whether these augmentation technologies can adapt to the evolution of AI.
\end{enumerate}

This study provides valuable insights into the potential of BO and BMI for augmenting human intelligence in an era of advancing AI.

\subsection{Paper Outline}

The remainder of this paper is organized as follows.
In section \ref{Features}, we explain the three features we focus on in this study.
In Section \ref{Modelling}, we model the BO, BMI, and hybrid systems.
In section \ref{Evaluations}, we evaluate BO, BMI, and hybrid systems using the three features.
In section \ref{Scenario}, we analyze how BO, BMI, and hybrid systems may be selected in an AI advancement scenario.
In section \ref{Discussion}, we discuss the results.
In section \ref{RelatedWork}, we review related work.
In section \ref{Conclusion}, we present the conclusion and future work.

\section{Augmentation Technologies Evaluating Features for User Decision} \label{Features}

As shown in Figure~\ref{fig::Overview}, in this paper, we examine three potential risks that users may be concerned about when considering intelligence augmentation by BO or BMI as a countermeasure to AI advancements.

BO and BMI technologies face many ethical, social, and technical challenges. For BMI, key issues include privacy violations, social inequality, loss of individuality, inauthenticity, undervaluation of achievements, and autonomy risks \cite{BMIReview, BMIConcern}. For BO, concerns arise around ethical oversight, potential consciousness, societal impact, and the moral value of advanced brain organoids \cite{BOReview, BOConcern}.

While these challenges are diverse, this study focuses on how potential risks and benefits of BO and BMI might specifically affect user decision-making. We assume that autonomy is an important factor in whether individuals choose to adopt these technologies \cite{IC1, IC2}.  Under this assumption, we have selected three critical dimensions that influence user decision-making, specifically processing capacity, self-identity, and the authenticity of consent.

\subsection{Processing Capacity}

The first dimension, processing capacity, is critical for understanding how effectively an augmentation technology can expand the user's cognitive abilities. For users, this dimension addresses the question ``How much can the technology enhance the user's capacity to process the information?''

As AI continues to advance at an exponential pace, augmentation technologies must provide scalable processing capabilities to keep up with the increasing complexity and volume of information \cite{Chalmers2016Singularity, Kurzweil2005Singularity}. Users considering these technologies need to think about the extent to which an approach can meet the demands of enhanced decision-making, problem-solving, and creativity, ensuring they remain competitive in an AI-advanced world.

\subsection{Identity Risk}

The second dimension, identity risk, focuses on the risks posed by augmentation technologies, particularly BO, to the user’s sense of self and individuality. For users, a critical question is ``Will this technology compromise my identity or disrupt my continuity of consciousness?''

BO augmentation may risk altering the user's identity as BO become larger \cite{BOReview, BOConcern}. 
For users, it is essential to evaluate these identity risks when they select the augmentation methods.

\subsection{Consent Authenticity Risk}

The third dimension, consent authenticity risk, addresses the potential for augmentation technologies, particularly BMI, to compromise a user’s autonomy. This dimension is critical for answering the question ``Will my decisions remain my own, free from undue influence by external systems? ''

BMI systems introduce external computational devices into decision-making processes, which raises concerns about the legitimacy and authenticity of consent \cite{BMIReview, BMIConcern}. Users need to understand whether these technologies allow them to maintain their autonomy and make decisions that reflect their true intentions.

\section{Modeling of Intelligence Augmentation Approaches} \label{Modelling}

\begin{figure}[t]
\begin{center}
\includegraphics[width=1\linewidth]{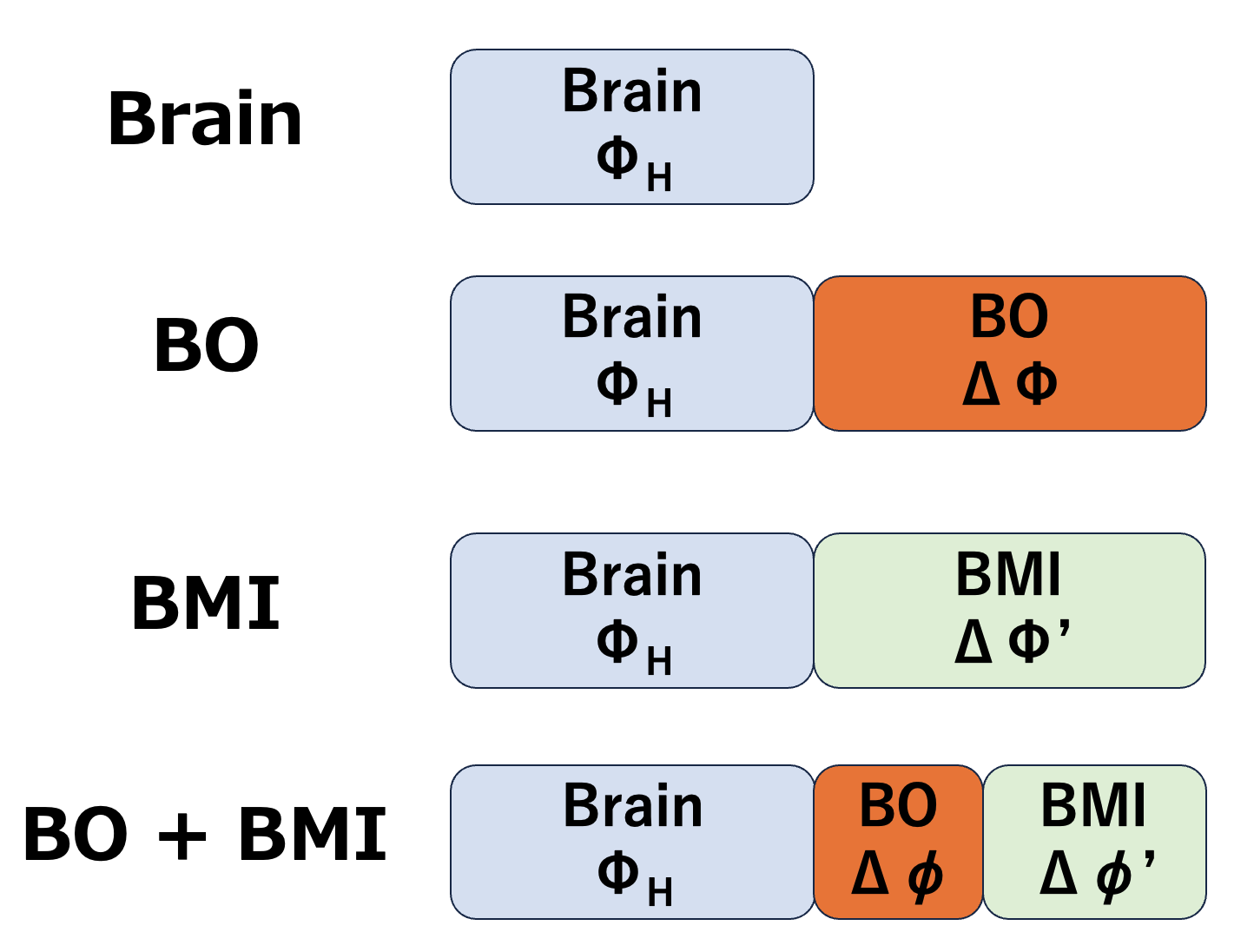}
\end{center}
\caption{Schematic representation of intelligence augmentation approaches.}
\label{fig::SchematicApproches}
\end{figure}

As shown in Figure~\ref{fig::SchematicApproches}, in this section, we classify and model the brain and intelligence augmentation patterns. 
These patterns include the natural brain, BO augmentation, BMI augmentation, and a hybrid approach combining BO and BMI.
Each pattern is characterized by the organization of information processing sites and information processing capabilities per site, as follows.

In the top of Figure~\ref{fig::SchematicApproches}, the natural brain capacity is represented as a brain with an information processing capability of $\Phi_H$.
This represents the baseline cognitive capacity without any external augmentation.

In the middle-top of Figure~\ref{fig::SchematicApproches}, the BO augmentation capacity is represented as $\Phi_H + \Delta\Phi$, where $\Phi_H$ is the natural brain's capacity, and $\Delta\Phi$ is the additional information processing capacity provided by BO. The parameter $\Delta\Phi$ reflects the increase in processing capacity introduced by the additional networks in BO, which can be enhanced by increasing the size or number of BO.

In the middle-bottom of Figure~\ref{fig::SchematicApproches}, the BMI augmentation capacity is represented as $\Phi_H + \Delta\Phi'$, where $\Phi_H$ is the natural brain's capacity, and $\Delta\Phi'$ is the additional information processing capacity provided by BMI. The parameter $\Delta\Phi'$ reflects the increaseed processing capacity introduced by the additional networks in BMI, which can be enhanced by increasing the size or number of BMI components.

In the bottom of Figure~\ref{fig::SchematicApproches}, the hybrid augmentation capacity is represented as $\Phi_H + \Delta\phi + \Delta\phi'$, where $\Delta\phi$ and $\Delta\phi'$ reflect the incremental information processing capacity provided by BO and BMI, respectively. These capacities can be enhanced by increasing the size or number of BO and BMI components.

\begin{table}[h]
\begin{tabular}{c|ccc}
                                                  & \begin{tabular}[c]{@{}c@{}}Processing\\ Capacity\end{tabular} & \begin{tabular}[c]{@{}c@{}}Identity\\ Risk\end{tabular}   & \begin{tabular}[c]{@{}c@{}}Consent\\ Authenticity\end{tabular} \\ \hline
Brain                                             & Limit                                                         & \begin{tabular}[c]{@{}c@{}}No\\ Risk\end{tabular}         & Limit                                                          \\ \hline
BO                                                & \begin{tabular}[c]{@{}c@{}}No\\ Limit\end{tabular}            & \begin{tabular}[c]{@{}c@{}}Uncontrolled\\ Risk\end{tabular}       & \begin{tabular}[c]{@{}c@{}}No\\ Limit\end{tabular}             \\ \hline
BMI                                               & \begin{tabular}[c]{@{}c@{}}No\\ Limit\end{tabular}            & \begin{tabular}[c]{@{}c@{}}No\\ Risk\end{tabular}         & Limit                                                          \\ \hline
\begin{tabular}[c]{@{}c@{}}BO\\ +BMI\end{tabular} & \begin{tabular}[c]{@{}c@{}}No\\ Limit\end{tabular}            & \begin{tabular}[c]{@{}c@{}}Controlled\\ Risk\end{tabular} & \begin{tabular}[c]{@{}c@{}}No\\ Limit\end{tabular}            
\end{tabular}
\caption{Comparison of intelligence augmentation approaches.}
\label{table::CmparisonTableh}
\end{table}

\section{Comparison of Augmentation Approaches} \label{Evaluations}
To evaluate the risks introduced by the augmentations shown in Figure~\ref{fig::Overview}, in this section, we examine information processing capacity, self-identity risk, and consent authenticity risk of different augmentation approaches.

\subsection{Information Processing Capability}

As shown in Table~\ref{table::CmparisonTableh}, we evaluate the information processing capacity of brain augmentation approaches.

The processing capacity of brain is represented as $\Phi_H$, which is inherently limited and can not be extended.

The processing capacity of BO augmentation is represented as $\Phi_H + \Delta\Phi$. The value of $\Delta\Phi$ can grow arbitrarily, depending on the size or number of BO.
Under the assumption of sufficiently advanced engineering and resource availability, there is effectively no upper limit to this growth.

The processing capacity of BMI augmentation is represented as $\Phi_H + \Delta\Phi'$. The value of $\Delta\Phi'$ can also be arbitrarily expanded by expanding or improving the machine interface.
Assuming sufficiently advanced technology and resources, this implies no effective upper limit.

The processing capacity of hybrid augmentation is represented as $\Phi_H + \Delta\phi + \Delta\phi'$. Both $\Delta\phi$ and $\Delta\phi'$ can be arbitrarily expanded, leading to no upper limit.

\subsection{Identity Preserving Capability}
As shown in Table~\ref{table::CmparisonTableh}, we evaluate the identity risks associated with various augmentation approaches.

Prior studies \cite{BOReview, BOConcern} suggest that large BO could influence the user’s sense of self-identity. We formalize this with the following assumption:

\textbf{Assumption}: Larger BO increase the self-identity risk.

Let $S_{BO}$ represent the size of the augmented BO, and let $f_{RISK}: \mathbb{R}_{\geq 0} \to \mathbb{R}_{\geq 0}$ be a function representing the magnitude of the risk associated with $S_{BO}$. The risk is expressed as $f_{RISK}(S_{BO})$
where $f_{RISK}$ satisfies the following conditions:
\begin{align*}
f_{RISK}(0) &= 0 \quad \text{(no risk when $S_{BO} = 0$)}. \\
f_{RISK}(S_{BO}) &> 0 \quad \text{for } S_{BO} > 0. \\
\frac{\mathrm{d}}{\mathrm{d}S_{BO}} f_{RISK}(S_{BO}) &> 0 \quad \text{(risk increases with size)}.
\end{align*}

In the case of BO augmentation, the risk is proportional to the size of $\Delta\Phi$ or $\Delta\phi$.
In contrast, BMI augmentation does not involve BO, and the risk is $f_{RISK}(0) = 0$.

\subsubsection{Natural Brain and BMI}
For the natural brain, there is no $\Delta\Phi$ or $\Delta\phi$, resulting in $S_{BO} = 0$, and the identity risk is $f_{RISK}(0) = 0$.

For BMI augmentation, there is no $\Delta\Phi$ or $\Delta\phi$, so $S_{BO} = 0$, and the identity risk is $f_{RISK}(0) = 0$.

In conclusion, as shown in Table~\ref{table::CmparisonTableh}, neither the natural brain nor BMI augmentation introduces identity risks by BO.

\subsubsection{BO and Hybrid Systems}

When using BO alone for augmentation, the identity risk is given by $f_{RISK}(\Delta\Phi)$.
In the case of hybrid systems, the identity risk is expressed as $f_{RISK}(\Delta\phi)$.

To compare hybrid systems and BO augmentation under the same level of processing requirements, we assume that the total processing capacity is equivalently augmented in both cases. Namely, we consider the following relationship:
\[
\Phi + \Delta\Phi = \Phi + \Delta\phi + \Delta\phi'.
\]

From this equation, we derive:
\[
\Delta\Phi = \Delta\phi + \Delta\phi' > \Delta\phi.
\]

Thus:
\[
f_{RISK}(\Delta\phi) < f_{RISK}(\Delta\Phi).
\]

In other words, because the hybrid approach distributes some of the capacity gain to the BMI (\(\Delta\phi'\)), the BO component (\(\Delta\phi\)) is smaller than in BO alone augmentation, leading to a lower identity risk. 
Namely, under equivalent performance requirements, a hybrid system achieves the same total processing capacity as BO alone augmentation but poses a lower and more controllable identity risk. 

In conclusion, as shown in Table~\ref{table::CmparisonTableh}, while BO alone can present an uncontrollable risk, the hybrid approach mitigates and regulates that identity risk effectively.

\subsection{Consent Authenticity Capability}
As shown in Table~\ref{table::CmparisonTableh}, we examine the consent authenticity risk for each augmentation approach.

BMI involves the processing of information by external machines during decision-making processes. This introduces the concern of the consent authenticity risk, as the influence of external systems may compromise the validity of users' consent.

We evaluate the limitations of consent authenticity in BMI augmentation under the following assumption:

\textbf{Assumption}: Any information processing that directly determines the validity of consent must be performed by the user's brain.

\subsubsection{Natural Brain}

For the natural brain, there are no concerns regarding consent authenticity, but the processing capability is limited to $\Phi$.
Therefore, as shown in Table~\ref{table::CmparisonTableh}, the natural brain's consent authenticity capability is limited.

\subsubsection{BO}
BO augmentation does not inherently pose authentic consent concerns, as it does not rely on BMI, and $\Phi + \Delta\Phi$ is not limited. Therefore, as shown in Table~\ref{table::CmparisonTableh}, BO's consent authenticity capability is unlimited.

\begin{figure*}[t]
\begin{center}
\includegraphics[width=1\linewidth]{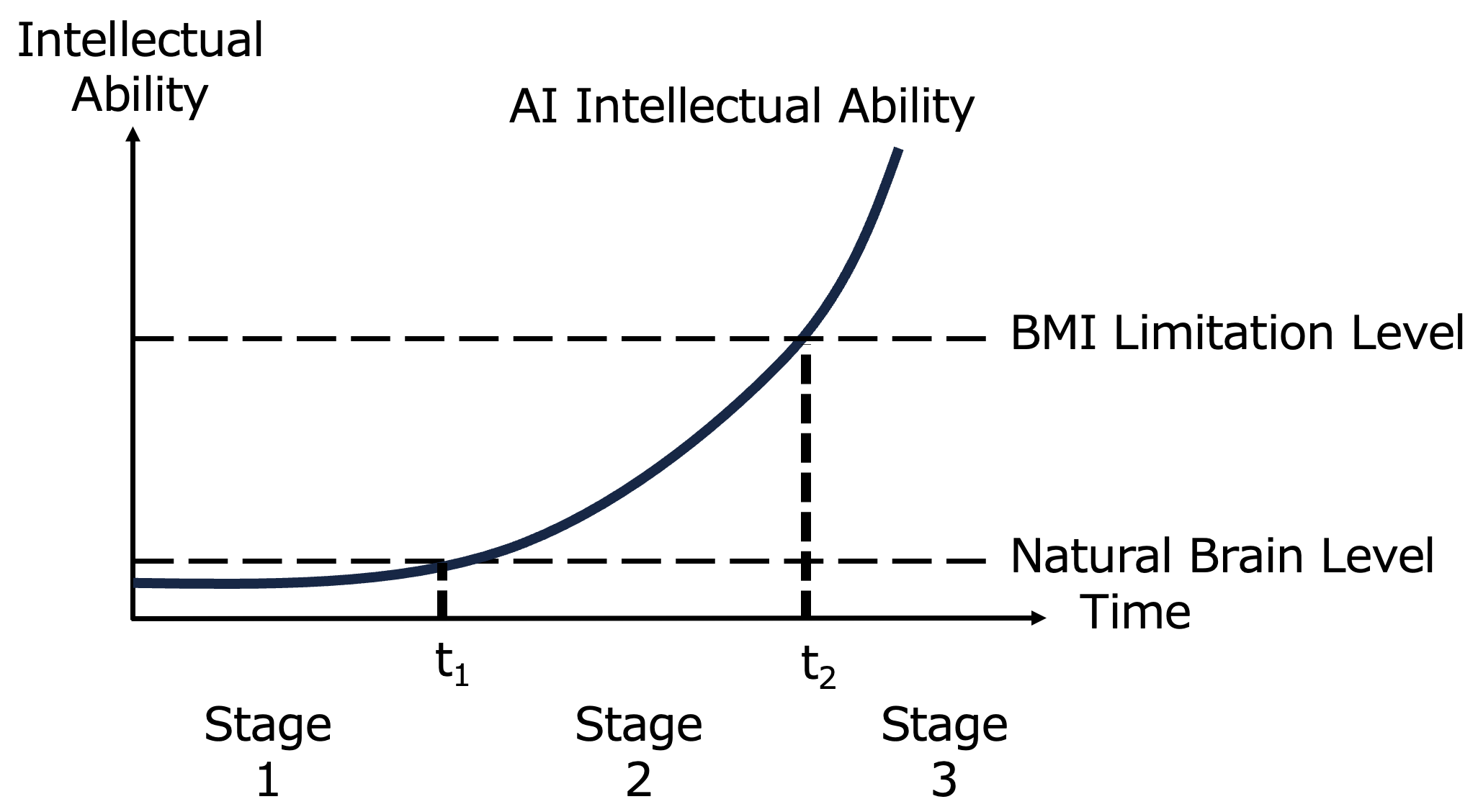}
\end{center}
\caption{Illustration of scenario analysis.}
\label{fig::Scenarios}
\end{figure*}

\subsubsection{BMI}

In BMI augmentation, the external machine ($\Delta\Phi'$) complements the brain's processing capacity ($\Phi_H$).

Let $I_{\text{BMI}}$ represent the amount of processing required per unit of time by the human for consent under BMI augmentation.
Since $I_{\text{BMI}}$ relates to consent authenticity, its crucial decision-making steps must be executed within the brain.

There are three methods for processing $I_{\text{BMI}}$ in BMI augmentation.
\begin{enumerate}
    \item Process $I_{\text{BMI}}$ using only external machines.
    \item Process $I_{\text{BMI}}$ exclusively within the brain.
    \item Process $I_{\text{BMI}}$ through the interaction between the brain and external machines.
\end{enumerate}

In the first case, where $I_{\text{BMI}}$ is processed using only external machines, consent processing can not be achieved.

In the second case, where $I_{\text{BMI}}$ is processed exclusively within the brain, consent processing is limited to the capacity of the natural brain ($\Phi_H$).

In the third case, $I_{\text{BMI}}$ is processed through the interaction between the brain and external machines. While the total processing capacity of the BMI system itself is not inherently limited due to the inclusion of the external machine ($\Delta\Phi'$), we have to investigate whether the ability to process consent remains constrained by the brain's natural capabilities ($\Phi_H$). This distinction is crucial for understanding whether authentic consent can be achieved under BMI augmentation.

When $I_{\text{BMI}}$ is processed through the interaction between the brain and external machines, some processing tasks can be offloaded to the external machine ($\Delta\Phi'$). However, the total processing capacity of the BMI system is ultimately constrained by the brain's ability to manage the information flow between itself and the external machine.

We denote $I_{\text{flow}}$ as the minimum amount of information required to be processed during such interactions between the brain and the BMI system.
Since $I_{\text{BMI}}$ is determined by both the processing performed by the brain ($\Phi_H$) and the amount of information exchanged between the brain and the external machine ($I_{\text{flow}}$), it can be expressed as:
\[
I_{\text{BMI}} = g_{\text{consent}}(\Phi_H, I_{\text{flow}}).
\]
where $g_{\text{consent}}$ is a function that is monotonically increasing with respect to both $\Phi_H$ and $I_{\text{flow}}$.

The information flow $I_{\text{flow}}$ is subject to the following constraint:
\[
I_{\text{flow}} \leq \Phi_H.
\]

Therefore:
\[
I_{\text{BMI}} = g_{\text{consent}}(\Phi_H, I_{\text{flow}}) \leq g_{\text{consent}}(\Phi_H, \Phi_H).
\]

In this context, $g_{\text{consent}}(\Phi_H, \Phi_H)$ is considered a constant for a given $\Phi_H$, representing the maximum processing capacity for consent under BMI augmentation.

This equation argues that regardless of the performance of the external machine ($\Delta\Phi'$), the overall processing capacity of the BMI system is ultimately dependent on the brain's ($\Phi_H$) capacity.

In conclusion, as shown in Table~\ref{table::CmparisonTableh}, BMI's consent authenticity capability is limited.

\subsubsection{Hybrid}

In hybrid systems, we examine whether there is an upper limit to consent authenticity.

We denote the total processing capacity related to authentic consent in hybrid systems as $I_{\text{Hybrid}}$. The value of $I_{\text{Hybrid}}$ is determined by the contributions from both the brain and the brain organoid ($\Phi_H + \Delta\phi$), as well as the minimum required information flow ($I_{\text{flow}}$) between the brain and the external machine.
This relationship can be expressed as:

\[
I_{\text{Hybrid}} = g_{\text{consent}}(\Phi_H + \Delta\phi, I_{\text{flow}}).
\]

The following constraint applies to $I_{\text{flow}}$:

\[
I_{\text{flow}} \leq \Phi_H + \Delta\phi.
\]

Thus, the total capacity can be rewritten as:

\[
I_{\text{Hybrid}} = g_{\text{consent}}(\Phi_H + \Delta\phi, I_{\text{flow}}) \leq g_{\text{consent}}(\Phi_H + \Delta\phi, \Phi_H + \Delta\phi).
\]

This indicates that $I_{\text{Hybrid}}$ increases as $\Phi_H + \Delta\phi$ increases, and since $\Delta\phi$ has no upper limit, $I_{\text{Hybrid}}$ is theoretically unlimited.

In conclusion, as shown in Table~\ref{table::CmparisonTableh}, the capacity for authentic consent in the hybrid system is unlimited.

\section{Scenario Analysis of Intelligence Augmentation Strategies in AI Development} \label{Scenario}

As shown in Figure~\ref{fig::Scenarios}, in this section, we analyze scenarios emphasizing the evolving relationship between humans and AI, as well as the strategic considerations for augmentation across three distinct stages.

We define AI's intellectual ability as $\Phi_\text{AI}(t)$, and $\Phi_\text{AI}(t)$ is assumed to increase exponentially over time, reflecting the rapid pace of innovation in AI development:
\[
\Phi_\text{AI}(t) = \Phi_\text{AI}(0) e^{kt}, \quad k > 0.
\]
where \(t\) is time, \(k\) is the exponential growth rate, and \(\Phi_\text{AI}(0)\) represents the AI's ability at \(t=0\). We take \(t=0\) to be an initial point where the AI’s capability has not yet surpassed that of a typical human.

Additionally, we assume that humans prioritize the preservation of personal identity and consent authenticity while adopting only the minimum required level of augmentation necessary to remain competitive with AI.

\subsection{Stage 1: Natural Brain Sufficient ($t < t_1$)}

At $t = t_1$, AI's intellectual ability reach parity with human intelligence:
\[
\Phi_\text{AI}(t_1) = \Phi_\text{H}.
\]

In this stage 1 ($t < t_1$), AI capabilities $\Phi_\text{AI}(t)$ have not yet surpassed baseline human intelligence $\Phi_\text{H}$, making intelligence augmentation unnecessary:
\[
 \Phi_\text{H} > \Phi_\text{AI}(t) \quad  (t<t_1).
\]

In Stage 1, natural human cognition remains sufficient to navigate the challenges posed by AI. This stage represents a period of relative equilibrium between human and artificial intelligence, where no augmentation, such as BO or BMI, is required.

\subsection{Stage 2: Augmentation Becomes Necessary ($t_1 \leq t < t_2$)}

At $t = t_2$, the capabilities of BMI augmentation reach their maximum potential, expressed as:
\[
\Phi_\text{AI}(t_2) = \max(\Phi_\text{H} + \Delta\phi').
\]

Here, $\max(\Phi_\text{H} + \Delta\phi')$ represents the upper consent authenticity limit of the augmentation capacity achievable through BMI as shown in Table \ref{table::CmparisonTableh}.

As time progresses beyond $t_1$, AI capabilities $\Phi_\text{AI}(t)$ surpass baseline human intelligence $\Phi_\text{H}$.
\[
\Phi_\text{AI}(t)  > \Phi_\text{H}   \quad (t_1 \leq t < t_2).
\]

Thus, necessitating augmentation to remain competitive as follows:
\[
\Phi_\text{H} + \Delta\Phi \geq \Phi_\text{AI}(t) > \Phi_\text{H},
\]
\[
\Phi_\text{H} + \Delta\Phi' \geq \Phi_\text{AI}(t) > \Phi_\text{H},
\]
\[
\Phi_\text{H} + \Delta\phi + \Delta\phi' \geq \Phi_\text{AI}(t) > \Phi_\text{H}.
\]
\[
   \quad (t_1 \leq t < t_2).
\]

In Stage 2, the natural brain alone is insufficient to compete with AI. Intelligence augmentation becomes essential and can be achieved through BMI add-ons ($\Delta\Phi'$), BO add-ons ($\Delta\Phi$), or a hybrid system that combines both ($\Delta\phi + \Delta\phi'$).

At this stage, the decision to use BMI or BO may depend on factors such as personal preferences, tolerance for risks, and the specific cognitive abilities that need enhancement. 

\subsection{Stage 3: BO or Hybrid as the Only Options ($t_2 \leq t$)}

When AI capabilities $\Phi_\text{AI}(t)$ continue to grow exponentially beyond $t_2$, BMI add-ons will eventually reach their limits, leaving BO or hybrid systems as the primary viable options for augmentation:
\[
\Phi_\text{AI}(t) > \max(\Phi_\text{H} + \Delta\Phi')   \quad (t > t_2).
\]

Thus, BO or hybrid are competitive as follows:
\[
\Phi_\text{H} + \Delta\Phi \geq \Phi_\text{AI}(t) > \Phi_\text{H} + \Delta\Phi',
\]
\[
\Phi_\text{H} + \Delta\phi + \Delta\phi' \geq \Phi_\text{AI}(t) > \Phi_\text{H} + \Delta\Phi',
\]
\[
   \quad (t > t_2).
\]

At this stage, augmenting with BO ($\Delta\Phi$) or a hybrid system ($\Delta\phi + \Delta\phi'$) becomes the only means for humans to counter AI advancements.

In Stage 3, the identity risk posed by BO and hybrid systems is particularly emphasized.
BO offers theoretically unlimited scalability through $\Delta\Phi$, enabling advanced cognitive augmentation. However, this scalability comes with significant risks, especially as the size of BO ($S_{BO}$) increase, potentially leading to identity risk. The risk $f_{\text{RISK}}(S_{BO})$ grows as the BO becomes larger, raising identity risk as shown in Table \ref{table::CmparisonTableh}.
On the other hand, hybrid system leverages the complementary strengths of BMI and BO while minimizing the risks associated with each. By integrating BMI with minimal BO augmentation, the hybrid approach maintains the processing advantages of BO while reducing the size ($S_{BO}$). 
This minimizes $f_{\text{RISK}}(S_{BO})$. Therefore, hybrid system minimizes the identity risk and consent authenticity risk shown in table \ref{table::CmparisonTableh}.
Namely, Hybrid augmentation emerges as a scalable and adaptive approach, ensuring both performance enhancement and identity risk minimization.

\section{Discussion} \label{Discussion}

\subsection{Result Evaluation}

As shown in Figure~\ref{fig::SchematicApproches}, we modeled BMI, BO, and hybrid systems as methods of intelligence augmentation to counter the rapid evolution of AI. As shown in Table \ref{table::CmparisonTableh}, we compared their augmentation capabilities and limitations. Moreover, as shown in Figure~\ref{fig::Scenarios}, we investigate the extent to which human intelligence augmentation can compete with AI advancements.
Our models indicate that while BMI is effective for enhancing human cognitive functions in the initial stages, BO become necessary for augmentations beyond a certain threshold. 

This result suggests a potential synergistic relationship between BO and BMI, where BMI may serve as a foundation for later integration with BO.
This study presents how the functional integration of transplanted BO and BMI balances the associated risks.
However, the integration of BO and BMI with each other is still in its early stages \cite{OIReview}, and the integration of BO and BMI into the human brain remains a challenge for future research.

\subsection{Limitation}

While this study focused on the limits of augmentation capabilities, we recognize that the speed of technological advancement is also crucial in countering AI's rapid evolution. As AI capabilities grow at an accelerating pace, augmentation technologies must evolve correspondingly to maintain parity. 

Future research should address the speed of technological advancements in intelligence augmentation, exploring models that incorporate factors like learning rates, technological adoption, and the dynamic interplay between humans and AI.

This study focuses on risks related to user autonomy, particularly self-identity and authenticity of consent, in the context of BO, BMI, and hybrid augmentation approaches. However, additional risks associated with BO and BMI need further investigation.

BO introduce a range of challenges beyond those modeled in this study, such as accountability, dependency, equitable access, and the broader implications for user self-determination \cite{BOReview, BOConcern}. 
How these risks can be incorporated into the proposed framework remains an open question for future research.

Similarly, BMI augmentation carries risks that extend beyond consent authenticity, such as privacy concerns, the undervaluation of BMI-assisted achievements, and the potential for external interference in decision-making \cite{BMIReview, BMIConcern}. These factors should also be evaluated to provide a more comprehensive understanding of BMI's ethical and practical implications.

For hybrid systems, which combine BO and BMI, the interplay between their respective risks creates additional complexities. Understanding how hybrid systems balance the potential benefits of expanded processing capacity with the compounded risks of self-identity disruption and autonomy challenges requires further theoretical and empirical exploration.

Future work should aim to expand the current framework to address these broader considerations, offering a more holistic approach to evaluating the ethical, social, and technical dimensions of augmentation technologies.

\section{Related Work} \label{RelatedWork}

\subsection{Augmentation Technologies in Competition with AI}

Approaches attempting to keep pace with the rapid evolution of AI through external technologies such as BMI and nanomachines face inherent limitations due to reliance on external augmentation \cite{Chalmers2016Singularity, Kurzweil2005Singularity}.
Discussions surrounding digital brain copying and mind uploading are also significant topics in intelligence augmentation to compete with AI. While issues of continuity of consciousness and identity in silicon-based mind uploading have been explored \cite{Chalmers2016Singularity}, perspectives on biological augmentation remain largely unaddressed.
Studies on enhancing the cognitive abilities of humans and AI \cite{Kurzweil2005Singularity} are well known, and dramatic cognitive improvements through human-AI integration have been discussed. However, such approaches do not consider the potential role of BO.

Our research addresses the challenges of add-on approaches in BMI, as discussed in mind uploading. Furthermore, our study introduces a novel perspective on biological intelligence augmentation using BO for internal biological enhancement, highlighting its potential to compete with AI.

\subsection{Enhancement Through BO}

BO is derived from stem cell-based brain tissue and has been widely applied in regenerative medicine and disease modeling, as well as in the fields of neuroscience and medicine. However, several issues have been raised \cite{BODevCephal}.
For example, the use of BO in creating chimera animals poses ethical concerns, and issues of identity have also been highlighted.

Our research envisions the use of BO as an intelligence augmentation device to counter the evolution of AI, with a particular focus on addressing the identity-related concerns of users. Furthermore, our study provides a unique comparison between BO and BMI from the perspective of AI advancements.

\subsection{Enhancement Through BMI}

BMI is a technology that directly connects the brain with computers to enhance cognitive and physical abilities, but several issues have been highlighted \cite{BMIReview, BMIConcern}.
Concerns such as hacking risks and limitations in processing capacity have been raised, with improvements in security and processing power being proposed to address these challenges.

Our study specifically addresses the limitations of BMI regarding consent authenticity.
Moreover, our research examines the potential of BMI from the perspective of AI advancements and provides a unique comparison with BO augmentation.

\subsection{Replacement and Add-on Augmentation}

Whether to replace or augment brain functions is a central theme in augmentation technologies \cite{BOReview, BOConcern, BMIReview, BMIConcern}. Numerous studies have explored silicon-based processors and neural prosthetics.
The ethical aspects of brain function replacement and augmentation raise technical and ethical challenges.
However, there are few detailed models that consider the limitations of augmentation from the perspective of AI alignment.

Our study focuses on the add-on types of BO and BMI, modeling them to determine which options for cognitive enhancement can effectively counter AI advancements.

\section{Conclusion} \label{Conclusion}

In this study, we modeled BMI, BO and a hybrid approach combining both as intelligence augmentation methods to counter AI's rapid evolution.

These methods were evaluated based on three aspects, namely the limits of processing capacity, identity risk, and consent authenticity risk.

Additionally, using our model, we conducted a scenario analysis to explore which intelligence augmentation methods users might choose in response to AI advancements.

We confirmed that the hybrid approach balances the identity risk associated with BO and the consent authenticity risk of BMI. 
The model also demonstrated that intelligence augmentation beyond a certain level can not be achieved with BMI alone.
Therefore, Achieving higher levels of augmentation requires either BO or a hybrid of BO and BMI.

For future work, we propose exploring the adaptability of the model to additional risks associated with BO and BMI. 
We also propose investigating its applicability when considering the pace of AI advancements.

\section*{Acknowledgments}

We would like to acknowledge the assistance of ChatGPT for English grammar and word choice adjustments.

The content has been fully reviewed, and the authors remain responsible.

\bibliography{aaai25}

\end{document}